\documentclass[linenumber, twocolumn]{aastex631}

\begin{document}


\title{A Significant Sudden Ionospheric Disturbance associated with Gamma-Ray Burst GRB 221009A}

\correspondingauthor{L. A. Hayes \& P.T. Gallagher}
\email{laura.hayes@esa.int, peter.gallagher@dias.ie}

\author[0000-0002-6835-2390]{Laura A. Hayes}
\affiliation{European Space Agency (ESA), European Space Research and Technology Centre (ESTEC), \\ Keplerlaan 1, 2201 AZ Noordwijk, The Netherlands}

\author[0000-0001-9745-0400]{Peter T. Gallagher}
\affiliation{Astronomy \& Astrophysics Section, DIAS Dunsink Observatory, Dublin Institute for Advanced Studies, Dublin, Ireland}

\begin{abstract}
We report a significant sudden ionospheric disturbance (SID) in the D--region of Earth's ionosphere ($\sim$60--100~km), which was associated with the massive $\gamma$--ray burst GRB 221009A on 2022 October 9. We identified the SID over northern Europe--a result of ionisation by X-- and $\gamma$--ray emission from the GRB--using very low frequency (VLF) radio waves as a probe of the D--region. These observations demonstrate that an extra--galactic GRB (z$\sim$0.151) can have a significant impact on the terrestrial atmosphere and illustrates that the Earth’s ionosphere can be used as a giant X-- and $\gamma$--ray detector. Indeed, these observations may provide an insight into the impacts of GRBs on the ionospheres of planets in our Solar System and beyond.         
\end{abstract}

\section{Introduction} \label{sec:intro}
On 2022 October 9, an extremely bright and long--duration GRB was detected by both Swift/Burst Alert Telescope (BAT) and Fermi/Gamma-ray Burst Monitor \citep[GBM;][]{meegan2009} instruments \citep[see][]{cir32632, cir32636}. With follow--up reports, it is now clear that this was the most energetic GRB ever observed, with photon energies reaching up to 18~TeV \citep{cir32677}. The GRB has now been reported in observations from many instruments, including the Spectrometer Telescope for Imaging X-rays \citep[STIX;][]{krucker2020spectrometer} on-board Solar Orbiter \citep{cir32611}. The source location of the GRB has been derived to be centred at RA~=~19h13m00s and Dec~=~ +19d48m34s and its red-shift estimated to be z~=~0.151 \citep{cir32388, cir32765}. Such energetic GRBs at close distances are extremely rare, and estimates put this as a once in a century (or longer) event \citep{cir32793}.

\section{The Ionosphere as a Giant Detector} \label{sec:observations}
The lowest lying region of Earth’s ionosphere, the D-region ($\sim$60--100~km in altitude), can be thought of as a giant high-energy detector of X-- and $\gamma$--rays from astronomical sources, including the Sun. As a partially ionised plasma, it responds in a characteristic way to ionising radiation from such sources. An example of such an ionising disturbance is that of solar flares, for which solar X--rays penetrate down to D--region altitudes, increasing electron densities to extents to cause significant space-weather impacts. 

The measurements of very low frequency (VLF; 3--30~kHz) radio waves that propagate in the waveguide formed between the Earth’s surface and the lower ionosphere provides a unique tool to remotely detect ionospheric disturbances. These measurements are often used to detect and study solar flares \citep[e.g.][]{mitra74, thomson2005, hayes2021solar}. In rare cases, extra-solar transients can also be detected in VLF measurements such as magnetar flares \citep{inan1999_mag}, SGRs \citep{Palit2018}, and even GRBs \citep{fishman1988observation}. However, typically these are identified on the night-side ionosphere. Here we report that the extremely bright GRB 221009A produced a sudden ionospheric disturbance so significant that it had a solar--flare sized signature on the day-side ionosphere.

\begin{figure} \label{figure_grb}
\centering
\includegraphics[width=0.45\textwidth]{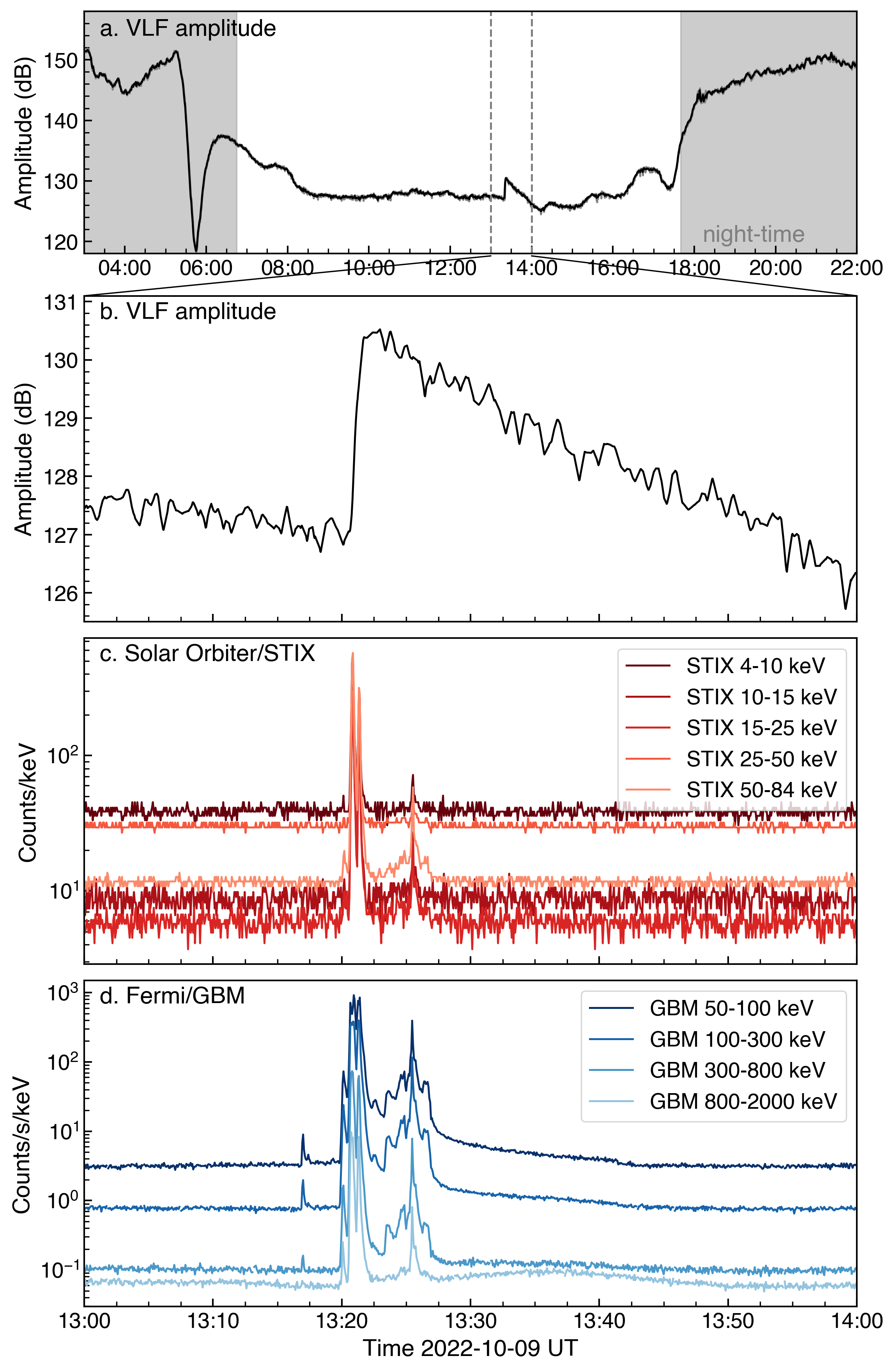}
\caption{ The VLF amplitude over the full day is shown in (a), with a zoom-in between the vertical lines shown in (b). The X-ray and $\gamma$-ray signatures of the GRB from STIX and the GBM (NAI-7) are shown in (c) and (d), respectively. The STIX time is adjusted for the light-travel time of the GRB between Earth and Solar Orbiter ($\sim$15~s) 
}
\end{figure}

\section{GRB and Ionospheric Observations} \label{sec:discussion}
The VLF measurements reported here are from a SuperSID monitor \citep{scherrer2008} that is operated at DIAS Dunsink Observatory in Dublin, Ireland (53.38$^{\circ}$~N, 6.33$^{\circ}$~W). The setup consists of a magnetic loop antenna (50~turns; 1.42~m side length). The monitor receives VLF signals from several worldwide transmitters, and is typically used for the study of X--ray solar flare effects on the D-region \citep[see][]{hayes2017pulsations, hayes2021solar}.


Figure~1~(a) shows the daily VLF amplitude measuring 23.4~kHz from the submarine communications transmitter (station ID: DHO38) located in Germany (53.09$^{\circ}$~N, 7.60$^{\circ}$~E). The grey shaded regions mark the night-time. The GRB can be identified between the vertical dashed lines. The zoomed--in plot of the VLF amplitude is shown in (b). The X--ray and $\gamma$--ray emissions from the GRB are shown in panels (c) and (d) from Solar Orbiter/STIX, and Fermi/GBM, respectively. The timing comparisons between the GRB high-energy signatures and the VLF response clearly indicates that the VLF peak is a result of the incident GRB fluxes. 


The VLF amplitude increase from the pre--GRB amplitude levels is $\sim$3.4~dB. To put this in perspective, based on statistical studies of solar flare effects this amplitude increase is equivalent to $\sim$C3--M1.0 GOES class X-ray flare. Furthermore, it is significant that this has been observed in the daytime-conditions. As discussed in \cite{raulin2014nighttime}, the night-time ionosphere is much more sensitive to external disturbances, when the solar radiation isn't dominating. And as such, the relative change in electron density required to produce a VLF amplitude change needs to be much larger during the day in order to measure it. This GRB was so large that it was detectable in the daytime observations. 

We also inspect the time--delay between the impulsive ionising radiation of the X-rays and $\gamma$-rays and the VLF response. For solar flares, there is a time--delay between the X--ray peak and the VLF response, typically around 2--3 minutes. Here we find a time-delay of $\sim$55~s. In terms of a recovery time, it takes almost 30 minutes to recover to pre-GRB levels following the impulsive ionisation burst.

\section{Discussion \& Future Work} \label{sec:discussion}
These observations show that GRB 221009A produced significant impacts on the ionisation of the lower ionosphere of our planet \citep[see also][]{cir32744, cir32745}. From our understanding, this is the strongest effect ever produced from a GRB and adds to the small handful of cases reported before. It is therefore clear that GRBs can have significant impacts on the ionospheres and indeed atmospheres of exoplanets, which could have implications for the evolution and stability of life on habitable exoplanets. 

The fact that this GRB was readily detected in the dayside ionosphere also has several interesting scientific implications, and provides an opportunity to study ionospheric chemistry from a new perspective. The short impulsive duration of the GRB acts as a delta-function input of ionisation into the ionosphere, and allows us to perform novel studies of the ionisation on short timescales and investigate the recombination recovery. Moving forward, we plan to investigate these observations in more detail, and undertake data--driven modelling of the ionosphere to understand the penetration depths altitudes of the GRB fluxes (which can be as low as 20~km), and investigate how the spectral components of the GRB influenced the response.


\software{matplotlib \citep{Hunter:2007}, pandas \citep{reback2020pandas}, sunpy \citep{sunpy_community2020}.}

\begin{acknowledgments}
L.A.H is supported by an ESA Research Fellowship. STIX is an instrument developed through international collaboration between Switzerland, Poland, France, Czech Republic, Germany, Austria, Ireland, and Italy. SuperSID was developed by Stanford University as part of International Heliophysical Year 2007. 
\end{acknowledgments}

\bibliography{rnaas.bib}{}
\bibliographystyle{aasjournal}

\end{document}